# Superconductivity of Cobalt in Thin Films

Nasrin Banu[1], M. Aslam[2], Arpita Paul[3], Sanjib Banik[4], S. Das[2], S. Datta[2], A. Roy[5], I. Das[4], G. Sheet[2], U. V. Waghmare[3] and B. N. Dev[1]*

[1]Department of Materials Science, Indian Association for the Cultivation of Science, 2A & 2B Raja S. C. Mullick Road, Kolkata 700032, India.
[2]Department of Physical Sciences, Indian Institute of Science Education and Research Mohali, Sector 81,S. A. S. Nagar, Manauli, PO 140306, India.
[3]Theoretical Sciences Unit, Jawaharlal Nehru Centre for Advanced Scientific Research, Jakkur, Bangalore 560 064, India.
[4]Condensed Matter Physics Division, Saha Institute of Nuclear Physics, 1/AF Bidhannagar, Kolkata 700064, India.
[5]Microelectronic Research Center, The University of Texas at Austin, 10100 Burnet Road, Bldg 160, MER 1.606J, Austin, Texas 78758, USA.

*msbnd@iacs.res.in, *bhupen.dev@gmail.com

**Abstract:** Due to competing long range ferromagnetic order, the transition metals Fe, Co and Ni are not superconductors at ambient pressure. While superconductivity was observed in a non-magnetic phase of Fe, stabilized under pressure, it is yet to be discovered in Co and Ni under any experimental conditions. Here, we report emergence of superconductivity in the recently discovered high-density nonmagnetic face centered cubic phase in Co thin films below a transition temperature ($T_c$) of ~5.4 K, as revealed in experiments based on point-contact spectroscopy and resistance, and four-probe measurements of resistance at ambient pressure. We confirm the non-magnetic nature of the dense *fcc* phase of Co within first-principles density functional theory, and show that its superconductivity below 5 K originates from anomalous softening of zone-boundary phonons and their enhanced coupling with electrons upon biaxial strain.

---------------------------------

**Main Text:** Normal metals possessing strong long-range magnetic order do not exhibit superconductivity (*1*). However, superconductivity can arise in some of these elements when subjected to pressure. Among the ferromagnetic transition metals, Fe is known to be non-magnetic at pressures above 10 GPa and exhibit superconductivity at temperatures below 2 K at pressures between 15 and 30 GPa (*2*). However, superconductivity has not been hitherto observed in Co and Ni under any conditions. It was predicted theoretically that, face-centered cubic (*fcc*) cobalt would lose its magnetic moment at a density of about 1.4 times the normal density (*3*). Recently, a high density non-magnetic (HDNM) cobalt with *fcc* structure was discovered in polycrystalline cobalt thin films of varied thickness, grown on silicon (*4, 5*). In direct lattice imaging by high resolution transmission electron microscopy, high-density *fcc* Co grains with (111) planar spacing of 1.83 Å, representing about 11% lattice (30 % volume) contraction, were observed (*4*) (see also Fig. S3 in (*6*)). Assuming an isotropic contraction, this corresponds to an *fcc* lattice parameter of 3.17 Å. Electronic structure of *fcc* Co with this lattice parameter, obtained here using first-principles calculations based on density functional theory with a local spin density approximation, confirms its non-magnetic ground state (see Fig. S5A in (*6*)) as observed experimentally.

Stimulated by the discovery of the HDNM - cobalt, we took up an exploration of superconductivity in HDNM Co, and report here its superconducting ground state that remains stable up to unusually high $T_c$ of about 9.5 K at ambient pressure. Superconductivity is revealed in the zero-bias conductance enhancement and its systematic evolution with magnetic field in mesoscopic point-contacts between normal non-superconducting elemental metals and the HDNM Co sample. An independent confirmation of superconductivity is obtained from the resistance drop at $T_c$ and the suppression of $T_c$ with applied magnetic field in both point contact measurement and standard four-probe measurement under ambient pressure.

A thin film (nominal thickness 25 nm) of Co was deposited on HF-etched clean Si(111) wafer substrate at room temperature by electron-beam evaporation. Exposure of the film to air has produced a thin (~ 2 nm) cobalt oxide on top of the film. The sample used in the present study is a part of what has been used for the discovery of HDNM cobalt (*4*). The Co film is polycrystalline and the sample structure, as obtained in Ref. (*4*), is CoO(2nm)/HDNM-Co(3.5nm)/normal-Co(18nm)/HDNM-Co(3.5nm)/Si(111), as shown schematically (Fig. S1) in Ref (*6*). We have used mesoscopic point contacts between normal metals (non-superconducting) like Ag and Pd and the HDNM Co sample. The physical contacts have been made on the top surface and microscopic cracks developed on the oxide surface made electrical contacts of mesoscopic dimension. We employed a lock-in based modulation technique to measure the differential conductance (*dI/dV*) of the point contacts as a function of applied DC bias voltage *V* (see Fig. S2, Ref (*6*)). These measurements have been carried out at different temperatures and different applied magnetic fields. In addition, temperature dependent resistance measurement was carried out at different applied magnetic fields. Existence of a superconducting phase of a material can be identified through such measurements (*7*). The structure of the sample along with experimental arrangement is available in Figs. S1- S2 of Ref. (*6*).

In the resistance versus temperature (R-T) data, obtained in point-contact measurements on a Ag-tip/HDNM-Co contact (Fig. 1A), resistance drops abruptly at 9.2 K. Upon application of external magnetic field, the resistance drop shifts to lower temperatures, and practically disappears above the field of 35 kG. This magnetic field dependence of resistance indicates a superconducting transition in the HDNM Co. The resistance below transition temperature does not go to zero for several reasons (*7*), among them is the contact resistance.

The *dI/dV* spectra, obtained in the point contact spectroscopy (PCS) measurements, of the HDNM-Co/Ag contact (Fig. 1B) show a sharp conductance peak at $V = 0$ at T=0.5 K. Usually a zero-bias conductance peak is observed for contacts between a normal (N) metal and a superconducting (S) metal. The observed conductance peak at $V = 0$ in Fig.1B is similar to that obtained for (N/S) point contacts between normal metal Cu and the superconducting Nb (*7*), and clearly indicates that the HDNM-Co is superconducting. The superconducting transition temperature ($T_c$) can be identified to be below 9.5 K, as the conductance peak vanishes at T = 9.5 K (Fig. 1B). The R-T plot (Fig. 1A) provides the value of $T_c$ to be 9.2 K. The *dI/dV* spectra (Fig. 1C) show a systematic change with increasing applied magnetic field (H) and the spectral features nearly disappear at 35 kG (Fig. 1C), an additional verification of the superconducting nature of HDNM-Co. A theoretical fit (Fig. 1D) to temperature dependent critical field $H_C(T) = H_C(0)[1- (T/T_C)^2]$ shows that the estimate of $H_C$ is consistent with the R-T and *dI/dV - V* measurements, although the data (Fig. 1D) deviate from the fitted curve. This is not unexpected because the sample contains nanoscale grains (*6*) of different sizes with a density variation and presumably with different strains. The dimension of contact on the sample and the number of such grains within the contact region may introduce some variation which is not clearly understood. As we will see later that the experiments with a Pd tip show a $T_c$ of 9.8 K, as measurements with a different tip involve a variation in contact size and different spots on the sample. The critical magnetic field at 0 K is 35 kG (Fig. 1D).(Figs. 1A and 1C).

It should be noted here that Andreev reflection could not be observed in our PCS experiment due to low mean free path of the film, which prohibited us from making point contacts in the ballistic regime.

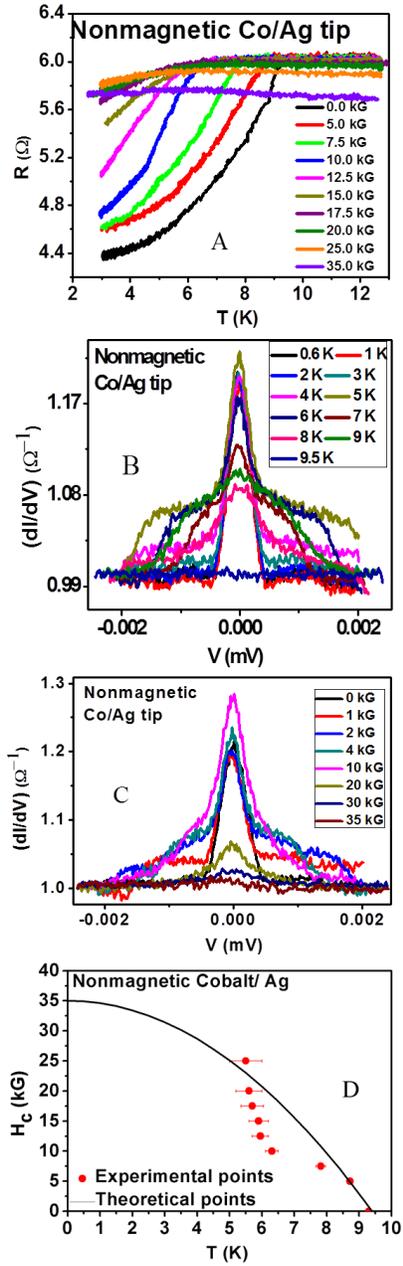

Fig. 1 A. R-T characteristics of the contact under different applied magnetic fields. B. Normalized conductance versus voltage characteristics of a nonmagnetic-Co/Ag contact at different temperatures at zero applied field. C. Normalized conductance versus voltage characteristics under different fields at T = 0.5 K. D. $H_c$–T plot shows the critical field, $H_c(0)$, to be about 35 kG.

We have repeated the experiments with a Pd (N) contact with the HDNM-Co sample confirming the superconducting transition of HDNM-Co. In zero external magnetic field, a sharp drop in resistance as a function of temperature (Fig 2A) occurs at T=9.8 K. The peak in conductance as a function of bias voltage (Fig. 2B) gradually diminishes with increasing temperature. A systematic change in the conductance vs voltage is observed with increasing applied field (Fig. 2C), and the disappearance of the spectral features occurs at 40 kG. In Fig. 2B and Fig. 2C, we notice the presence of conductance dips symmetric about $V = 0$, in addition to the conductance peak at $V = 0$. These features have also been observed in Ag(N)/Pb(S) N/S contacts (7), and our observations on Pd/HDNM-Co contacts clearly identify the superconducting state of HDNM-Co. The *dI/dV* spectral features in Fig. 2B do not disappear even at 9.3 K, implying that the $T_c$ is above this temperature.

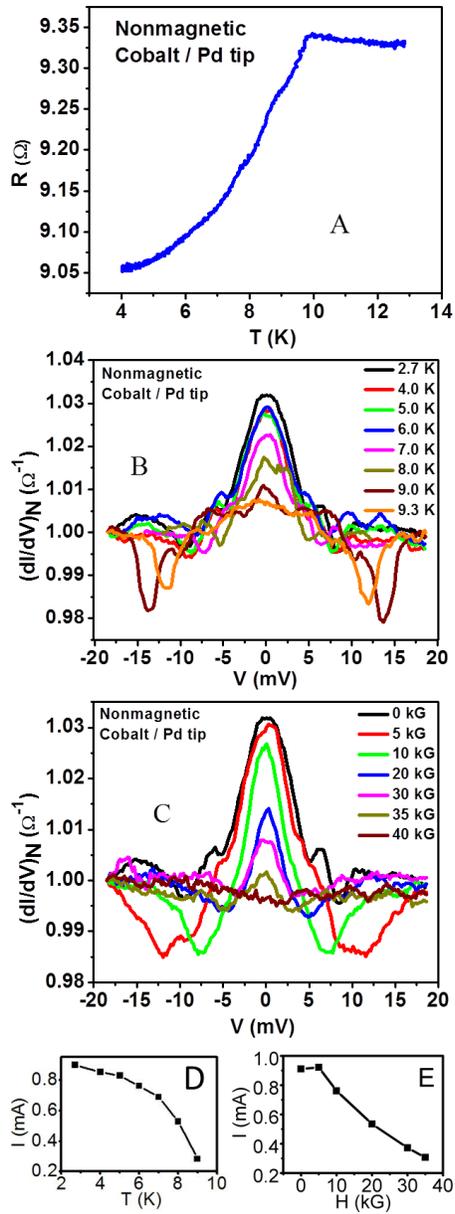

Fig. 2 A. R-T characteristics of a HDNM-Co/Pd contact at zero field. B. Normalized conductance - voltage characteristics of the contact at different temperatures at zero applied field. C. Normalized conductance-voltage characteristics under different fields at T = 2.7 K. D. Critical current - temperature derived from the positions of the symmetrical dips in Fig. 2B. E. Critical current - applied magnetic field derived from the symmetrical dips in Fig. 2C.

The symmetrical dips in the *dI/dV* spectra around *V*=0 carry information about the critical current. The value of the voltage at the dip position divided by the normal state resistance gives the critical current. (Figs. 2D and 2E).

We observe some variation in the results for the Ag tip and the Pd tip, for example in the width of the peak and the value of $T_c$. This is not unexpected as there is a dependence on the junction-type, and the junction size which include the tip material and the type of contact (*7-9*). As our nonmagnetic superconducting Co layers are polycrystalline with relatively small grains (typically ~ 10 nm) (*4*) (Fig. S3 in (*6*)), depending on the contact size and the region of the film where the contact has been made, there can be some variation in these results. Grain boundary regions with normal density (see Fig. S3) are expected to be normal metal. Inter-grain Josephson tunnelling may also be partly responsible for the observed differences (*10*).

The T-dependent resistance obtained from the four-probe measurements under ambient pressure (see Fig. 3A, with the inset showing two transition temperatures under vanishing magnetic field) shows that $T_{C2}$ (3.7 K) is related to superconductivity of In in the $In_xAg_y$ contact material. This is also observed in control experiments, where the sample has no superconductivity (Fig. S5B). Measurements at fields (500 Oe or larger) above the critical field do not show this transition. Our estimate of the superconducting transition in the HDNM Co is $T_{C1}$ =5.4 K, which reduces with applied field as indicated by arrows in Fig. 3A. The reason for resistance below $T_C$ not going to zero is explained in SM [6]. Here too, we find some deviation of the data from the theoretical fit to T-dependent $H_c$ (Fig. 3B).

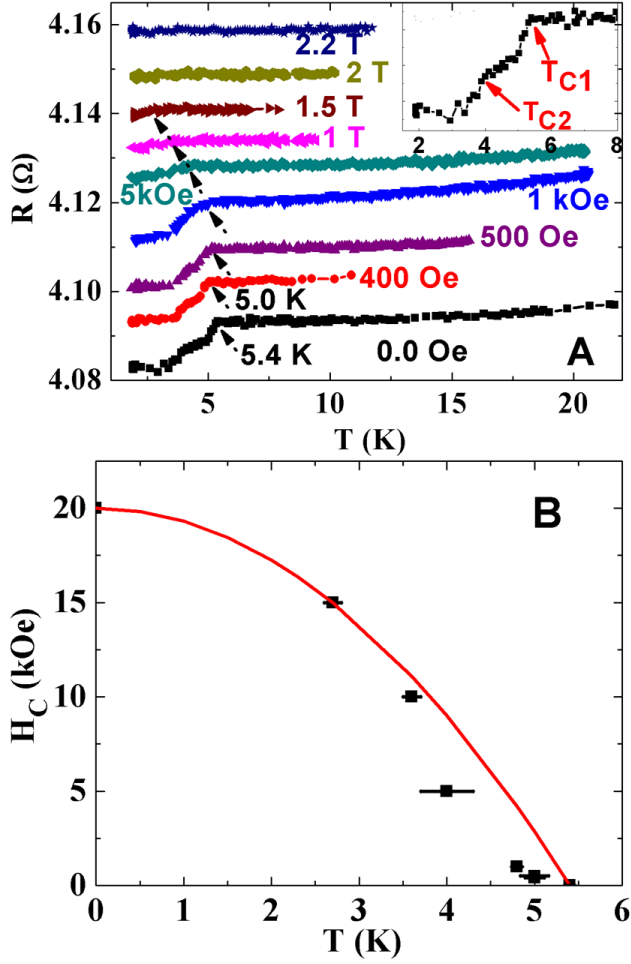

Fig. 3 A. R-T plots for various applied magnetic fields. See details in the text. The data for different fields are shown with a vertical offset for clarity. B. Critical field ($H_C$) vs temperature plot, obtained from A.

We estimated superconducting $T_C$ of nonmagnetic Co ($a_{exp}(fcc)$ = 3.17 Å) within the BCS theory using first-principles Eliashberg function in Mcmillan's equation (*11-13*). From the peaks in Eliashberg function (details in S6B (*6*)), we find that the phonon modes at X ($\omega$=333 cm$^{-1}$ and 433 cm$^{-1}$), W ($\omega$=335 cm$^{-1}$ and 385 cm$^{-1}$) and L ($\omega$=236 cm$^{-1}$) points of the Brillouin zone couple strongly with electrons. Our estimate of the dimensionless electron-phonon coupling of the HDNM cubic phase of Co ($\lambda$) is 0.32, and the consequent superconducting $T_C$ of nonmagnetic Co is 0.3 K, which is much smaller than its experimental value ($T_C$(exp) = 9.2 K - 9.8 K from point-contact and 4.4 K – 5.4 K from four-probe measurement), and it is due to weak electron-phonon coupling (small $\lambda$). It should be mentioned here that loss of magnetism in bulk Co under high pressure has been reported (*14*); however, no investigation was made regarding superconductivity.

To connect closer with the HDNM *fcc* phase of Co in the thin films studied here, we determined the effects of volume preserving anisotropic strain in the HDNM *fcc* structure that is expected to occur in a film. We used a body centered tetragonal unit cell (Fig. 4A) containing two Co atoms, strained biaxially ($e_{x'x'}= e_{y'y'}=e$, $e_{zz}= -2e$) so as to preserve the atomic volume of the observed HDNM *fcc* structure. From the phonon spectra of Co calculated as a function of *e*, we find anomalous softening of phonon modes (Fig. 4B, see Fig. S7) at N and X points ($\Delta\omega \approx$ 35-50 cm$^{-1}$) (*6*), as a result of their strong coupling with electronic states. Concurrently, the electron-phonon coupling constant ($\lambda$) increases sharply (Fig. 4C) with *e*, and consequently results in strongly nonlinear enhancement in superconducting $T_C$ (Fig. 4C) with strain near *e* ~ 0.05.

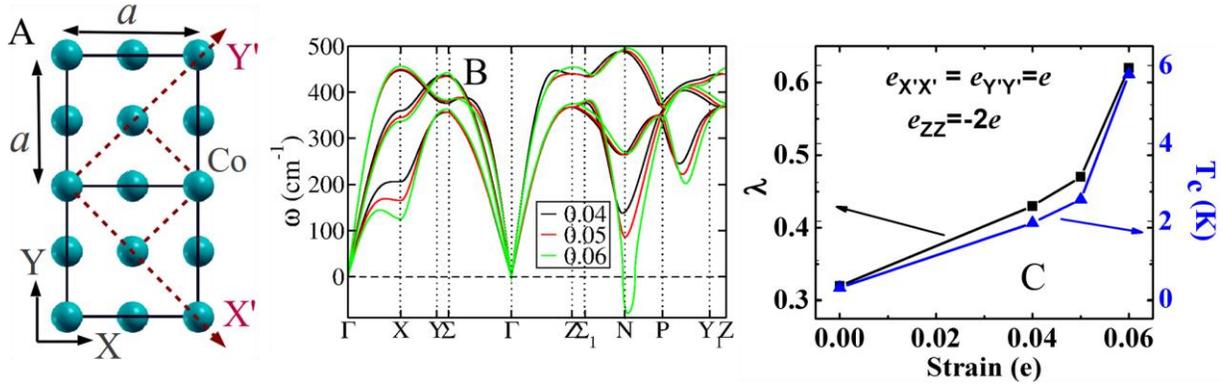

Fig. 4 A. The *fcc* crystal structure of Co, with a unit cell enclosed by red (dashed) lines representing the body centered tetragonal unit cell used as a reference while applying volume preserving biaxial strain (*e*). B. Phonon dispersion curves for *e* = 0.4, 0.5 and 0.6 showing anomalous softening of phonon modes at X and N points of the Brillouin zone. C. Electron-phonon coupling constant ($\lambda$) and superconducting transition temperature ($T_C$) as a function of strain (*e*).

Superconducting $T_C$ exceeds 5 K (Fig. 4C) at *e*=0.06 ($\lambda$=0.62), which is comparable to the observed $T_C$ ( ~ 5 to 9.5 K). This significant enhancement in superconducting $T_C$ of Co is governed by (a) enhanced electron-phonon coupling, accompanied by (b) unusual softening of phonon modes at N and X points with strain *e*, and (c) decrease in the electronic density of states at the Fermi level (see Table S1 in (*6*)). A strong dependence of the electron-phonon coupling ($\lambda$) on volume preserving strain *e* is partly due to the nesting between the electronic states in the neighbourhood of the Fermi energy.

We have observed superconductivity in a Co thin film containing high-density nonmagnetic (HDNM) Co. The superconducting state has been identified via observations of (i) field dependent transition in R-T, (ii) zero-bias conductance peak and its field evolution and (iii) conductance dips symmetric about V = 0 and their field evolution. The Co film is polycrystalline and the HDNM Co is in the form of nanoscale grains. There is a density, and possibly strain, distribution among the grains. That is why measurements of superconducting transition temperature on different spots on the sample show little variation in results ($T_C$: 9.2-9.8 K). Four-probe measurements have provided somewhat smaller value of $T_C$ (4.4 – 5.4 K). In point contact measurement, generation of some nonhydrostatic pressure on the sample at the point contact cannot be avoided. This might be the reason for the higher value of $T_C$ compared to the four-probe measurement. Our first-principles calculations confirm the nonmagnetic nature of Co (in the dense *fcc* phase) and show that a volume conserving strain in the HDNM *fcc* Co results in a remarkable enhancement in its superconducting $T_C$ ( > 5 K) due to anomalous softening of zone boundary phonons and their enhanced coupling with electrons.

**Acknowledgments**

We acknowledge the help given by Hema C. P. Movva in sample preparation. The work was partially supported by the DAE IBIQuS project (DAE OM No. 6/12/2009/BARC/R&D-I/50, Dated 01.4.2009). Nasrin Banu is supported by CSIR fellowship (09/080(0765)/2011-EMR-I). UVW acknowledges support from a JC Bose National Fellowship.

# Additional Supplementary Materials

**Materials and Methods**

**1. Thin film growth and sample structure**

A thin cobalt film of 25 nm nominal thickness was deposited on piranha cleaned, HF-etched Si(111) substrate (a 100 mm dia wafer) in high vacuum by electron-beam evaporation method. Then the cobalt film was taken out of the vacuum chamber. The exposure of the film to air led to surface oxidation. Several pieces were cut from this wafer sample for various experiments. X-ray reflectivity (XRR) experiment, which has a high depth resolution of ~1 Å, was carried out to obtain the density depth profile of the sample. This has provided the sample structure, as shown in Fig. S1, which shows two high density (HD) Co layers just below the CoO layer and just above the Si(111) substrate. The central part of the Co film has the usual density of Co. We have carried out polarized neutron reflectivity (PNR) experiment, which corroborate this layer structure. In addition PNR shows that the HD cobalt layers are nonmagnetic (NM) (*4*). The HD NM Co layers are shown in Fig. S1. Transmission electron microscopy (TEM) experiments – both planar and cross-sectional – have shown that the Co film is polycrystalline and the normal

Co at the middle of the structure is hexagonal close packed (hcp), whereas the HDNM Co is face-centered cubic (*fcc*) (*4*). The circuit diagram for the point contact measurements in shown in Fig. S2.

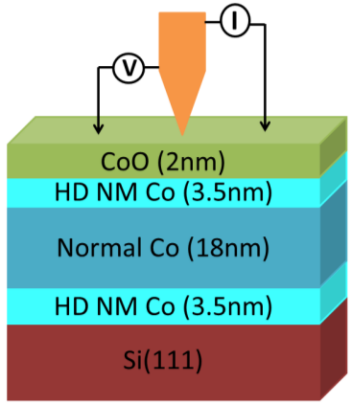

**Fig. S1** The layer structure in the sample, as obtained from XRR and PNR analysis in Ref. (*4*). The schematic of the point contact measurements is also shown.

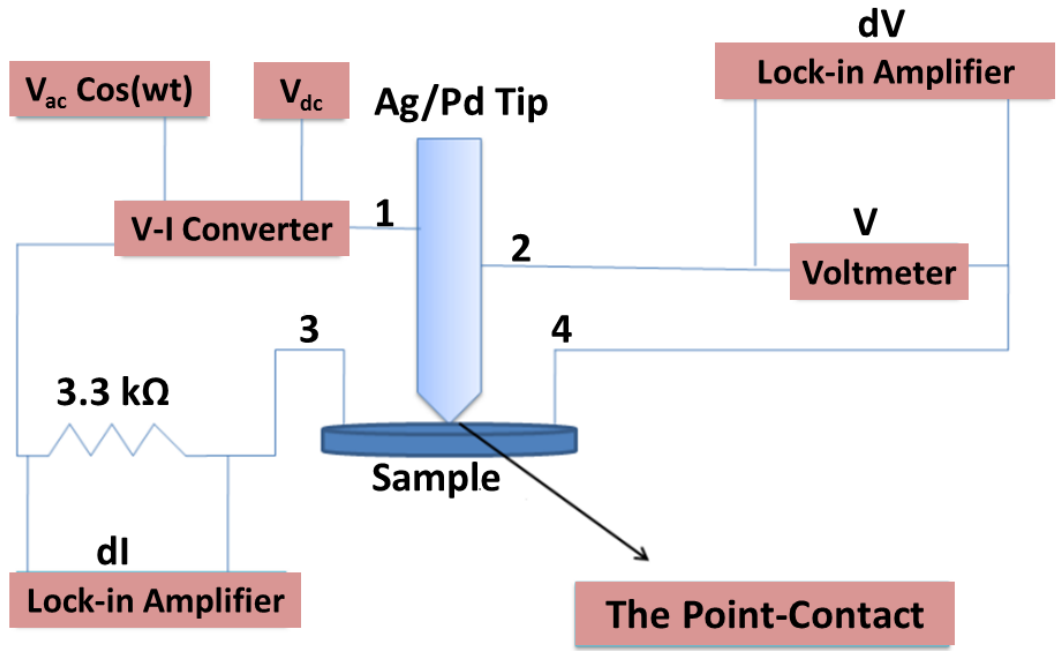

**Fig. S2** The circuit for the point contact measurements (*15*).

XRR and PNR experiments provide the value of the density at each depth, with sub-nanometer depth resolution, averaged over the whole sample area. This does not show the lateral variation of the density among the polycrystalline grains. This is seen in Fig. S3. The mechanism of how compressive strain, giving rise to higher density, is formed in polycrystalline thin films is explained in Ref. (*16*). During thin film deposition atoms can diffuse into grain boundaries and consequently the grains undergo compression. A planar TEM image (Fig. S3) from our sample shows a high density grain and its grain-boundary regions (*4*). The dark grain has *fcc* structure as seen in the fast Fourier transform (FFT) pattern in (b) from the boxed region '1' in (a). The high resolution TEM image (c) from this region shows a *fcc* (111) planar spacing of 1.83 Å. Images (d, e) from the grain boundary regions '2'and '3' shows the *fcc* (111) planar spacing of 2.06 Å just outside the high density grain. This is like normal *fcc* Co. The planar spacing of 1.83Å within the high density grain represents a 11% lattice (or ~ 30% volume) contraction. Considering an isotropic lattice contraction, this would give a *fcc* lattice parameter of 3.17Å.

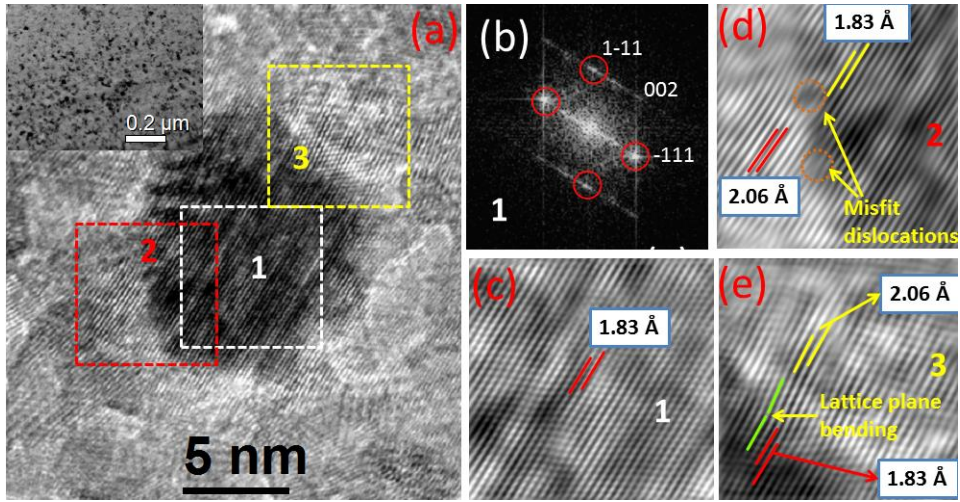

**Fig. S3** Plan view TEM images from the sample. A grain with a lattice contraction of 11%, leading to high density, is highlighted. See text for details. (From. Ref. *4*).

## 2. Ruling out any possibility of Nb in the sample

The observed value of $T_c$ for the nonmagnetic Co is strikingly similar to that of bulk Nb (9.26 K). This is apparently a coincidence. In thin films of Nb the value of $T_c$ depends on the film thickness and is much smaller compared to bulk Nb. For Nb films on Si, $T_c$ is 9.21 K for a film thickness of 214 nm and it reduces to 5.35 K for a film thickness of 22.5 nm (*17*). For Nb films grown on fused quartz, measurements carried out for a film thickness of 6.5 nm showed a $T_c$ value of 4.36 K. In our case the thickness of the nonmagnetic Co is only about 3 nm. The total thickness of the Co film is only about 25 nm. So, even if the whole film were Nb, the Tc would be much smaller than 9.2 K, the smaller observed value for our Co film. Moreover, the observed critical field ($H_c$) for our Co system is a few times larger than that for bulk Nb. So, our observed superconductivity has nothing to do with Nb. We demonstrate the absence of Nb in our sample by Rutherford backscattering spectrometry (RBS) experiment and simulation results (Fig. S4) (*6*). RBS experiments (*18, 19*) were carried out using 1 MeV He+ incident ions and by detecting the energy distribution of the scattered ions at a backscattering angle of 165°.As shown in the simulated RBS spectrum (red), the presence of Nb, equivalent to even one atomic layer, either on the Co film or distributed within the film, would have been observed in the experiment. For this case a Nb peak, as shown in the inset of Fig. S4 (red), would have been observed, as the signal would be much above the background. However, the observed spectrum (black) shows only the background and no peak.

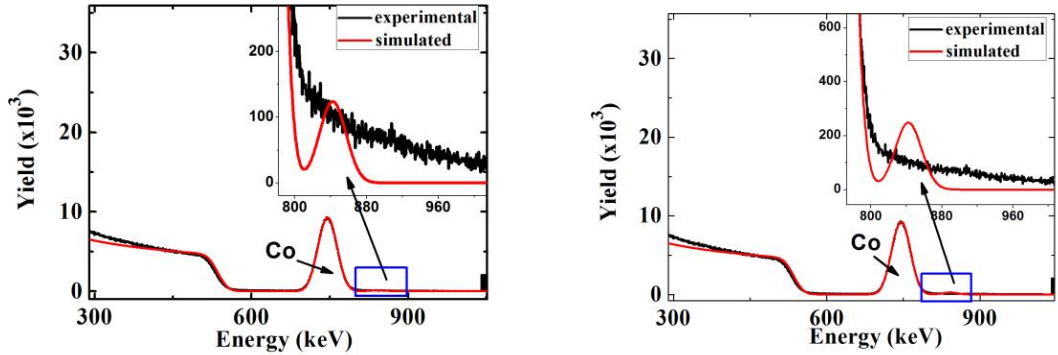

**Fig. S4** A. Measured RBS spectrum from the Co film on Si sample (black) and the simulated spectrum for only one atomic layer of Nb on Co (or an equivalent number of Nb atoms distributed in the Co film) (red). The black curve is hidden behind the red one where the simulation has the best match with the data. B. Same as in A, but for two atomic layers of Nb. The blue boxed region, shown magnified in the inset, in both A and B is where the Nb peak would be expected if there were any Nb in the sample.

### 3. Four-probe resistance measurement under ambient pressure

For the four-probe measurements, contacts have been made on the top surface of the sample using an $In_xAg_y$ contact material. (Contacts made with Ag paint did not survive at low temperature). This contact material was made from 70 wt. percent of In and 30 wt. percent of Ag by arc melting. Dimensions of the contacts are about 1 mm and the separation between successive contacts is 1 mm. In Fig. 3 we notice two superconducting transitions. However, the resistance below the transitions never goes to zero. For our sample the resistance is not expected to go to zero below the superconducting transition. As we notice from Fig. S3, around the high density grains, i.e., in the grain boundary regions the lattice parameter is not contracted and hence the density is normal. These regions are expected to be like normal metal around the HDNM superconducting grains even below the superconducting transition temperature. Additionally, the thin CoO layer as well as the normal Co below the HDNM Co layer contributes to the measured resistance. The schematic of the four-probe measurement and the sample structure are shown in Fig. 5A. In the four-probe experiment a 10 μA current was passed through the outer connectors.

For a reference sample, a Cu/Co multilayer, the resistance (R) versus temperature (T) plot at zero applied magnetic field is shown in Fig. S5B. It shows the superconducting transition of the contact material at 3.7 K, close to the superconducting transition temperature of pure In.

As the Cu/Co material between the contacts is normal metal, zero resistance below the transition is not expected.

Measurements on the HDNM Co samples at zero field shows a superconducting transition at 5.4 K, in addition to the one at 3.7 K (Fig. 3A). At applied magnetic fields of 500 Oe and above the contact material becomes normal metal and only the superconducting transition of HDNM Co survives. Eventually HDNM Co becomes a normal metal at the critical field of 20 kOe.

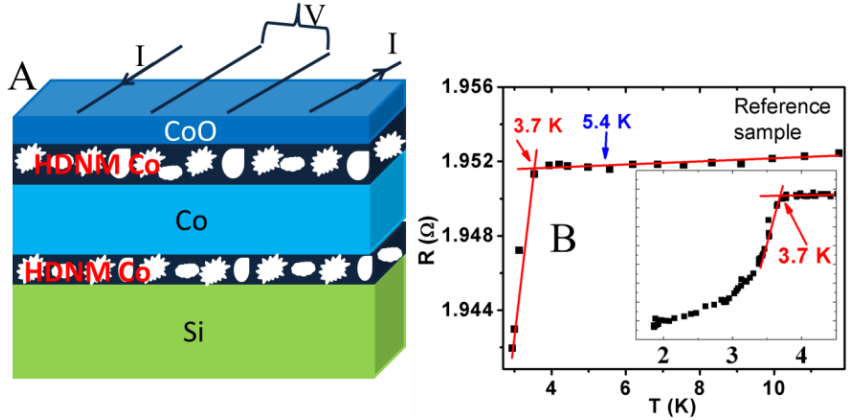

**Fig. S5** A. The schematic of the four-probe measurement on the HDNM Co sample. B. R vs T plot for a Cu/Co multilayer sample where the same $In_xAg_y$ contact material, as for the HDNM Co sample, has been used. The inset shows the details around the transition point. The position of the transition temperature (5.4 K) of the HDNM Co sample is marked.

From Fig. S5B we notice that the resistance does not go to zero below the transition temperature. Here the superconductivity comes from the contacts only and the material between the contacts is normal metal (thin Cu/Co multilayer).

## 3. Computational details

We determined electronic structure of *fcc* Co within first-principles density functional theory as implemented in QUANTUM ESPRESSO package (*20*), treating spin dependent exchange-correlation energy within a local density approximation (LDA) in Perdew-Zunger parametrized functional (*21*). We employed ultrasoft pseudopotentials (*22*) to model the interactions of ionic cores with valence electrons, and used an energy cutoff 30 Ry to truncate the plane wave basis in expansion of wavefunctions and 240 Ry in the representation of charge density. Brillouin Zone integrations (*fcc* lattice) were sampled on 8×8×8 uniform meshes of k-points.

Dynamical matrices and phonons were determined at wave vectors (q's) on 4×4×4 meshes in the Brillouin Zone using density functional perturbation theory (QUANTUM ESPRESSO (*19*) implementation). Dynamical matrices at other q-points were obtained from Fourier interpolation of those obtained at q's on 4×4×4 mesh. In addition, we used 64×64×64 meshes of k-points (electrons) and 4×4×4 meshes of q-points (phonons) to compute electron-phonon coupling matrix elements. Superconducting $T_C$ was estimated within the BCS theory (*11*) (Bardeen, Cooper and Schriefer) using Mcmillan's equation (*12, 13*),

$$T_c = \frac{\omega_{log}}{1.2} exp\left[\frac{-1.04(1+\lambda)}{\lambda(1-0.62\mu^*)-\mu^*}\right] \qquad (1)$$

where λ is the dimensionless electron-phonon coupling constant and μ* is the Coulomb pseudopotential (*13*), taken in our calculations as 0.1. $\omega_{log}$ is analogous to Debye frequency ($\Theta_D$), and expressed as (*12, 13*),

$$\omega_{log} = exp\left[\frac{2}{\lambda}\int \frac{\alpha^2 F(\omega)\log\omega}{\omega}d\omega\right] \qquad (2)$$

where ω's are the frequencies of phonon modes. $\alpha^2 F(\omega)$ is the Eliashberg spectral function.
We used experimental lattice parameters of Co ($a$ = 3.17 Å) in the *fcc* structure in our simulations.

## 4. Electronic structure and phonon dispersion

Electronic structure and phonon dispersion results are shown in Fig. S6. Fig. S6A shows the electronic density of states (DOS) for up-spin and down-spin states for a *fcc* Co with lattice parameter $a$ = 3.17 Å, corresponding to the observed high density phase of Co. The symmetrc up-spin and down-spin DOS indicates loss of magnetism. As the primitive *fcc* cell contains one Co atom, there are three phonon modes and all of which are acoustic (see Fig. 6B). Absence of the phonon modes with imaginary frequencies confirms the local stability of *fcc* structure of Co. Peaks in the Eliashberg spectral function (see Fig. 6B) show that phonon modes at X (ω=333 cm$^{-1}$ and 433 cm$^{-1}$), W (ω=335 cm$^{-1}$ and 385 cm$^{-1}$) and L (ω=236 cm$^{-1}$) points couple strongly with electrons.

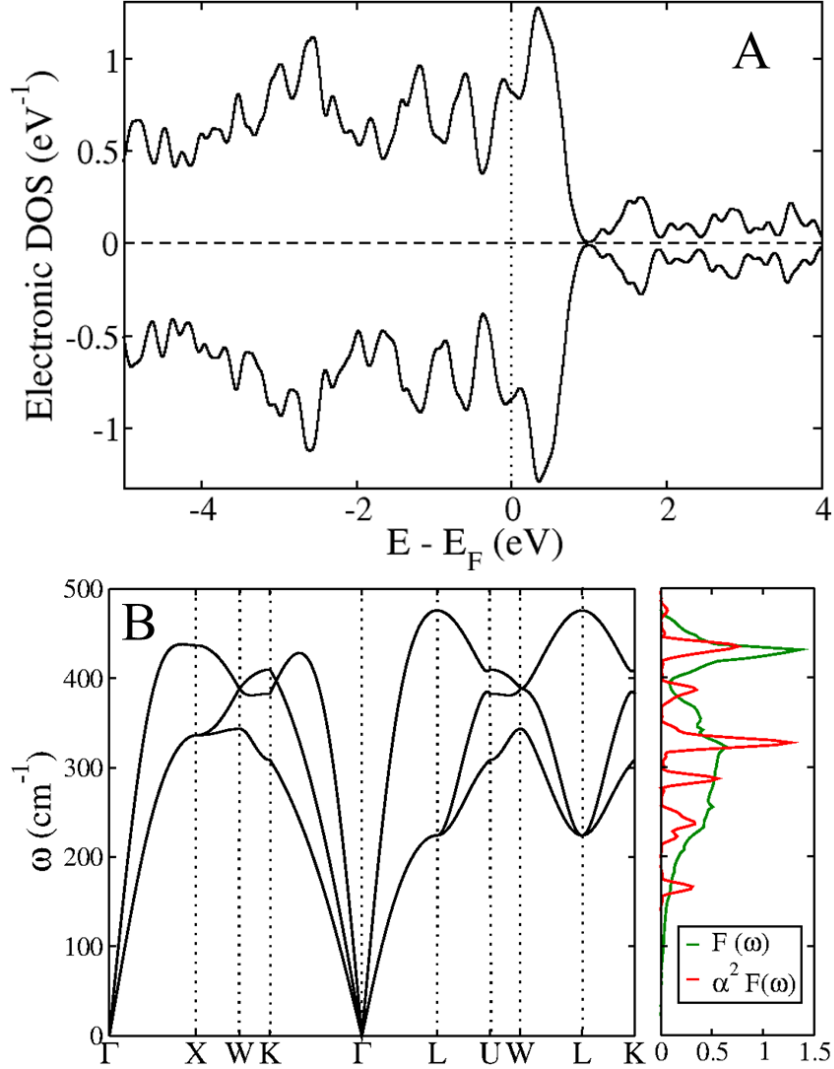

**Fig S**6 A. Electronic density of states of Co in the *fcc* structure. B. Phonon dispersion, phonon density of states ($F(\omega)$) and Eliashberg spectral function ($\alpha^2 F(\omega)$) of *fcc* Co.

### A. Effect of strain

To investigate the effect of strain on superconducting $T_C$, we consider body centered tetragonal unit cell (see Fig. 4A) containing two Co atoms. As (010) surface of tetragonal structure ((110) plane of *fcc* structure) of Co is hinted by experiments, we apply uniaxial compressive strain ($e_{x'x'} = 0.0$, $e_{y'y'} = -0.02$, $e_{zz} = 0.0$) along Y' axis ([110] direction of the *fcc* structure). Superconducting $T_C$ of Co increases slightly to 0.48 K from 0.29 K as λ (=0.33, see Table S1) does not change significantly after applying uniaxial strain. We now consider volume preserving strain ($e_{x'x'} = 0.04$, $e_{y'y'} = -0.08$, $e_{zz} = 0.04$) in the tetragonal unit cell. We find that λ enhances to 0.37 (from 0.32, see Table S1) and the superconducting $T_C$ rises to 1.05 K (from 0.29 K). As the value of λ is small (0.33 and 0.37) in both the cases, our predicted superconducting $T_C$ is still smaller compared to its experimental value (9.2 K).

TABLE S1: Estimated electron-phonon coupling constant ($\lambda$), density of states at the Fermi level ($N(\varepsilon_F)$) and superconducting $T_C$ of Co as a function of strain.

| $e_{x'x'}, e_{y'y'}, e_{zz}$ | $\lambda$ | $N(\varepsilon_F)$ (States/eV) | $T_C$ (K) |
|---|---|---|---|
| (0.0,0.0,0.0) | 0.32 | 0.73 | 0.29 |
| (0.0,-0.02,0.0) | 0.33 | 0.71 | 0.48 |
| (0.04,-0.08,0.04) | 0.37 | 0.83 | 1.05 |
| (0.04,0.04,-0.08) | 0.43 | 0.82 | 1.96 |
| (0.05,0.05,-0.1) | 0.47 | 0.75 | 2.56 |
| (0.06,0.06,-0.12) | 0.62 | 0.74 | 5.76 |

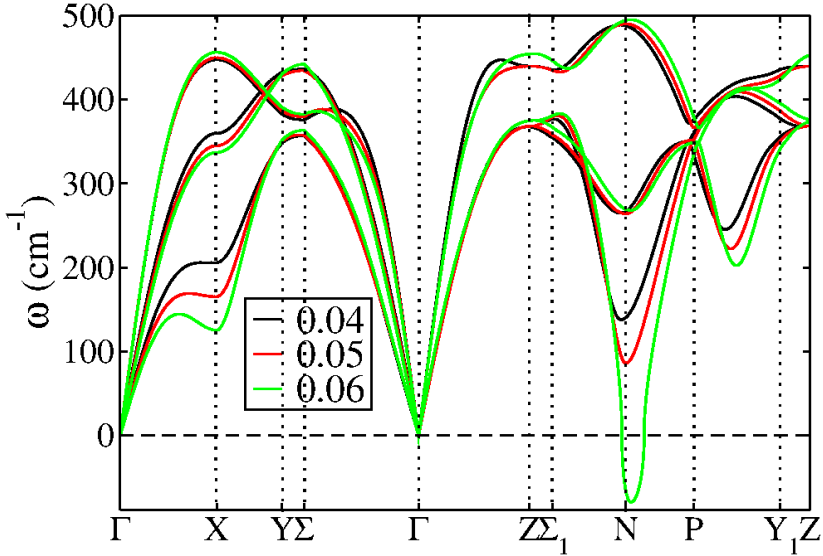

**FIG. S7**: Phonon dispersion (obtained considering primitive body centered tetragonal cell which contains one Co atom) of Co as a function of volume preserving strain e ($e_{x'x'} = e_{y'y'} = e$, $e_{zz} = -2e$).

We notice that a small magnetic moment per Co atom arises for higher values of strain ($e > 0.04$). For $e = 0.06$ ($e_{x'x'} = e_{y'y'} = e$, $e_{zz} = -2e$), we estimated $\lambda$ by ignoring the weak structural instability ($\omega = i85$ cm$^{-1}$, see Fig. S7) present at N point of the Brillouin Zone. Enhancement in the strength of coupling of phonon modes (at N and X points) with electronic states after introducing strain (*e*) on the tetragonal cell can be explained as follows. Occupied down-spin bands at Γ and N points with energies of -1eV (-0.65 eV for up-spin band) and -0.05 eV (-0.38 eV for up-spin band) are almost at *e* =0 .06 (see Fig. S8A and Fig. S8B) and lead to high electronic density of states at those energies. Phonons with wave vectors $q_X$ $(=(0.5,0.5,0.0)\frac{2\sqrt{2}\pi}{a})$ and $q_N$ $(=(0.5,0.0,0.43)\frac{2\sqrt{2}\pi}{a})$ scatter down-spin electrons in the occupied states to the down-spin unoccupied states (see Fig. S8B) at X (with energy of 0.7 eV) and Γ (with energy of 0.2 eV) points. Similarly, $q_X$ and $q_N$ scatter up-spin electrons in the occupied states (at Γ and N, see Fig. S8A) to the up-spin unoccupied states at X (with energy of 0.69 eV) and Z (with energy of 0.18 eV) points respectively. The electron-phonon coupling constant (λ) is expressed as (*23, 24*),

$$\lambda = \sum_{qv} \frac{2}{\hbar \omega_{qv} N(\epsilon_F)} \sum_k \sum_{ij} \left| g_{k+q,k}^{qv,ij} \right|^2 \times \delta(\epsilon_{k+q,i} - \epsilon_F)\delta(\epsilon_{k,j} - \epsilon_F) \qquad (3)$$

where $N(\varepsilon_F)$ is the density of states at the Fermi level, and $g_{k+q,k}^{qv,ij}$'s are the electron-phonon coupling matrix elements. $\omega_{qv}$ is the frequency of the $v^{th}$ phonon mode at wave vector q.

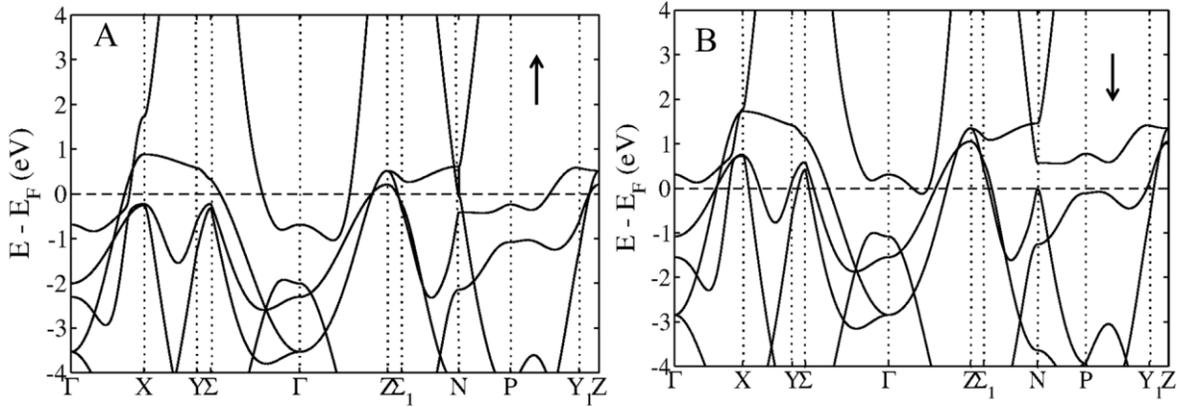

**Fig. S8**: A and B represent spin resolved electronic structure of Co for *e* = 0.06 ($e_{x'x'} = e_{y'y'} = 0.06$, $e_{zz} = -0.12$).

Phonon modes with low frequencies (phonons at N and X points) contribute much more to λ (Eq. 3) compared to high-frequency modes.

It should be noted here that the theoretical calculations are for materials of uniform density. In this case a parabolic relationship between $H_C$ and T is expected for BCS superconductivity. In the experiment the sample is composed of high density grains with some variation in density of the grains These high density grains are also surrounded by normal metal. The sample is as if superconducting grains are embedded in a normal metal. This is likely to be the reason for the deviation of the $H_C$ – T relationship from parabolicity.